\documentclass[twocolumn,showpacs,amsmath,amssymb]{revtex4}

\usepackage{amsmath}
\usepackage{amsfonts}
\usepackage{amssymb}
\usepackage{graphicx}
\usepackage{dcolumn}
\usepackage{bm}
\usepackage{subfigure}
\usepackage{CJK}

\graphicspath{{Figures/}}
\begin{document}


\title{Effect of Ridge-Ridge Interactions in Crumpled Thin Sheets}


\author{Shiuan-Fan Liou$^1$, Chun-Chao Lo$^1$, Ming-Han Chou$^1$, Pai-Yi Hsiao$^2$, and Tzay-Ming Hong$^{1,3}$}
\affiliation{$^1$Department of Physics, National Tsing Hua University, Hsinchu 30013, Taiwan, Republic of China\\$^2$Department of Engineering and System Science, National Tsing Hua University, Hsinchu 30013, Taiwan, Republic of China\\$^3$Center for Fundamental Science Research, National Tsing Hua University, Hsinchu 30013, Taiwan, Republic of China}

\date{\today}

\begin{abstract}
We study whether and how the energy scalings based on the single-ridge approximation are revised in an actual crumpled sheet; namely, in the presence of ridge-ridge interactions. Molecular Dynamics Simulation is employed for this purpose. In order to improve the data quality, modifications are introduced to the common protocol. As crumpling proceeds, we find that the average storing energy changes from being proportional to one-third of the ridge length to a linear relation, while the ratio of bending and stretching energies decreases from 5 to 2. The discrepancy between previous simulations and experiments on the material-dependence for the power-law exponent is resolved. We further determine the averaged ridge length to scale linearly with the crumpled ball size $R$, the ridge number as $1/R^2$, and the average storing energy per unit ridge length as $1/R^{2.364\sim 2.487}$. These results are consistent with the mean-field predictions. Finally, we extend the existent simulations to the high-pressure region for completeness, and verify the existence of a new scaling relation that is more general than the familiar power law at covering the whole density range.
\end{abstract}


\maketitle

	Although crumpling is ubiquitous and simple to enact, intense researches to understand its complexities only began in the last thirty years or so. It has not only become relevant to cutting-edge technologies like utilizing crumpled graphene sheets\cite{graphene} to harvest energies by converting motion into electricity, but is of interest to the general phenomenon of condensation of many outstanding problems in physics\cite{wood}. However, some properties of crumpling remain unresolved. For instance, as the developable cones\cite{cerda,cerda2,cerda3,cerda4} increase in number and form the familiar network\cite{ridge,ridge2,witten} of ridges and vertices, scientists still do not know how they collaborate to produce stunningly simple power laws between the crumpled ball size $R$ and crumpling force $F$\cite{kantor,kantor2,seung,matan,vliegenthart,balankin,lin,tallinen1,tallinen2,bai}, $R\sim F^{-\alpha}$ and for the occurring frequency\cite{noise,noise2,noise3} of different noise intensities. One challenge is to understand how the crumpled sheet constructs spontaneously a highly porous and yet robust structure\cite{tomography,tomography2,tomography3}. Proper theoretical tools are also in demand to tackle the complex many-body interactions among these deformations, especially the self-avoidance that plays an important role\cite{vliegenthart} at creating the glass-like interior after a series of highly non-equilibrium processes\cite{aharoni}, similar to the random packing in a golf ball basket or salt jar.  
	
	 In view of the theoretic inability, Molecular Dynamics Simulation becomes a powerful tool to obtain information such as the three-dimensional distribution of ridges and facets before unfolding, energy storage in each ridge, and how the increase of crumpled ball density affects these quantities. We are interested in the effect of ridge-ridge interactions on the energy scaling predicted by the single-ridge approximation\cite{ridge,witten,wood}. Attentions are also paid to resolve major disagreements between previous simulations\cite{vliegenthart,tallinen2} and experiments\cite{lin,bai}.  
In the mean time, we push the simulation to a more-time-consuming region of large densities to verify the existence of a new scaling law\cite{bai}. Finally, we study how the ridge number, averaged ridge length, and energy cost per unit length evolve with the crumpled ball density, and compare these relations with the mean-field predictions. 

	Our simulations follow a protocol which models a thin sheet by a triangular lattice 
with bond length $r=1$\cite{vliegenthart,tallinen2,seung}. Bending and stretching moduli\cite{seung}, $k_B$ and $k_S$, are imposed on a two-dimensional circular membrane with radius $R_0$=130 lattice sites, which composes of $N_{\rm bead}=62143$ beads. Simulation on a larger system of $R_0$=500 sites, for test, has been checked not to alter our conclusions. The WCA\cite{wca} Lennard-Jones potential is used to insure that no bead can penetrate each other. Crumpling force is simulated by a collapsing impenetrable spherical wall that clothes the membrane. In order to compare with the experiments, a realistic value of 1/3 is assigned to the Poisson ratio and the relation $k_B/k_S=3h^2/32$ is used to constrain $k_S$ with each choice of $k_B$. The bead diameter is held constant, $h=0.9$.  
We also include plasticity by halving the magnitude of $k_B$ beyond a yield angle of $10^{\rm o}$ and requiring the strain to relax with the original $k_B$.

\begin{figure}[!h]
  \centering
  \includegraphics[width=7cm]{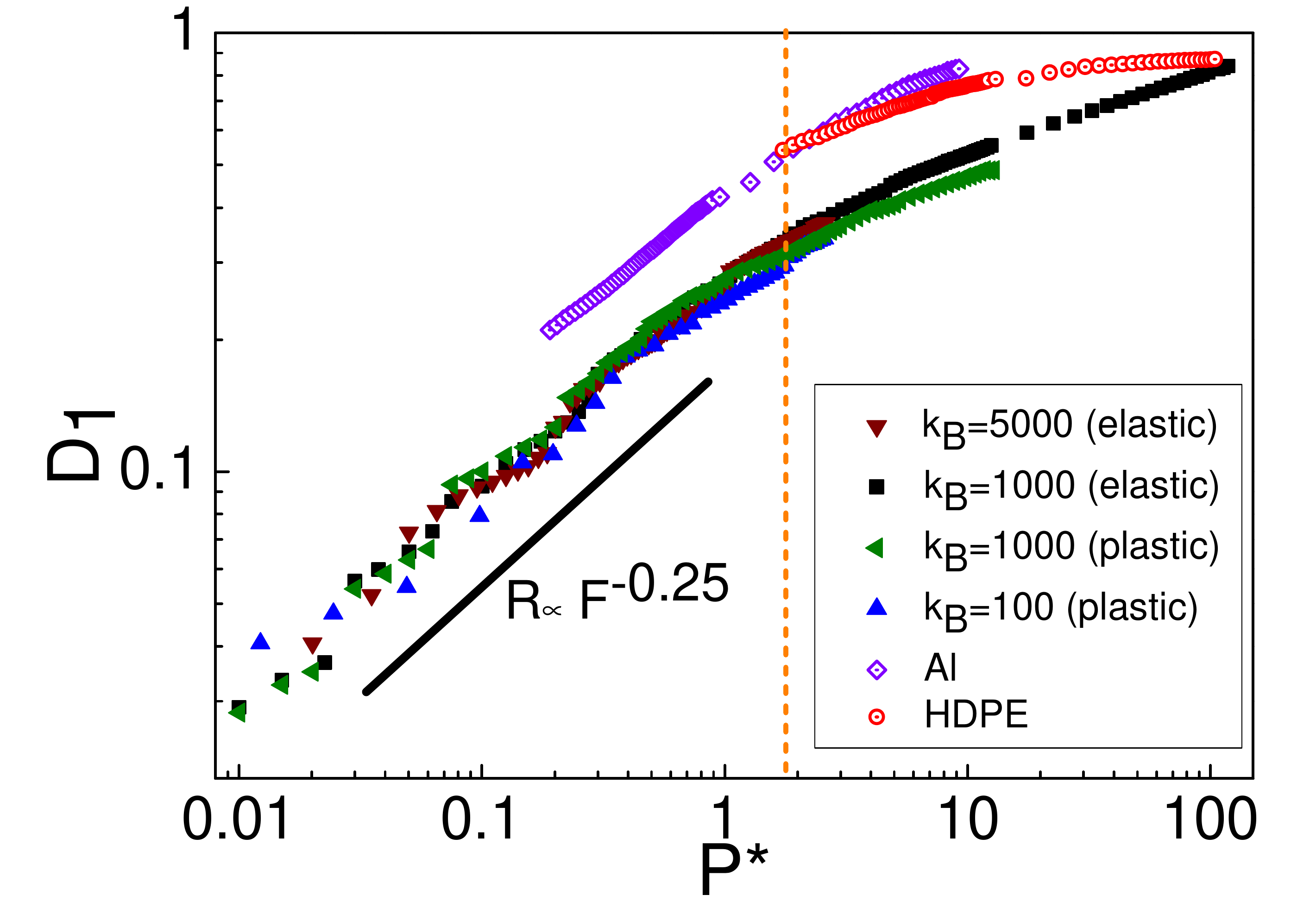} 
  \caption
  {Dimensionless density $D_1$ vs. dimensionless pressure $P^\star$ for circular elastic and plastoelastic sheets of different $k_B$. Experimental data\cite{bai} for Al and HDPE have been superimposed on the plot for comparison. The power-law region is marked on the left-hand side of the vertical dashed line. 
}\label{fig_P_Density} 
\end{figure}  

We first verify that our simulation reproduces the correct mechanical response.
The data plotted in Fig.~\ref{fig_P_Density} show how the (dimensionless) density of the crumpling sheet $D_1\equiv N_{\rm bead}[4\pi (h/2)^3/3]/[4\pi R_{\rm wall}^3/3]$ varies with the (dimensionless) external pressure $P^\star \equiv (P/Y)(R_0/h)$ where $R_{\rm wall}$ is the radius of the collapsing wall and $Y$ is the Young's modulus. The reason why we show $D_1$ vs. $P^\star$ here, instead of $R$ vs. $P$ (or $F$), is that 
the presentation collapses the simulation data obtained from different $k_B$ onto a master curve, irrespective of whether plasticity is included.
The results reveal a general scaling relation across the whole range of $P$, which can be reduced to a power law $D_1\sim P^{\star 3\alpha/(1+2\alpha )}$ at small density. We find that the exponent $\alpha$ is 0.25, in good agreement with simulations obtained by other groups~\cite{vliegenthart,tallinen2}.
However, discrepancy is observed when we superimpose experimental data~\cite{bai} on the same figure for comparison. The densities for Al and HDPE are significantly larger than the simulation ones.

\begin{figure}[!h]
\mbox{
\begin{minipage}[t]{0.5\linewidth}
\centering
\includegraphics[width=4.2cm]{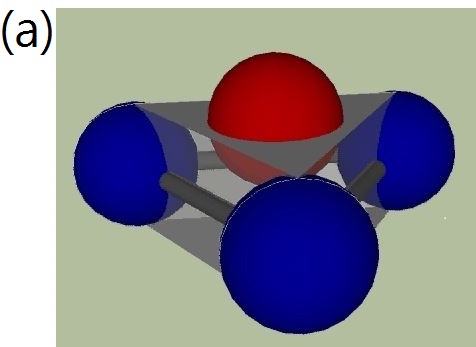}
\label{ball_new}
\end{minipage}
\begin{minipage}[t]{0.5\linewidth}
\centering
\includegraphics[width=3.5cm]{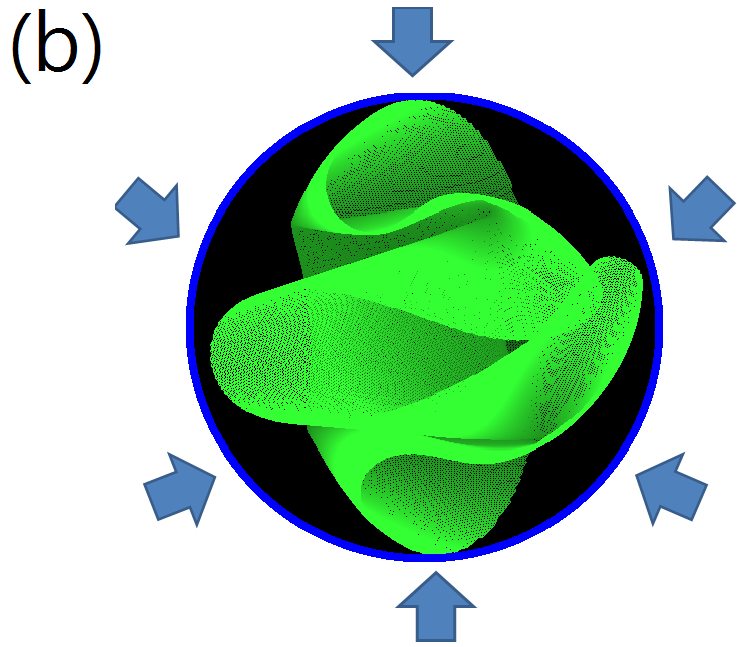}
\label{fig_mod2_new_b}
\end{minipage}
}
\mbox{
\begin{minipage}[t]{0.5\linewidth}
\hspace*{-1.5cm}\includegraphics[width=7cm]{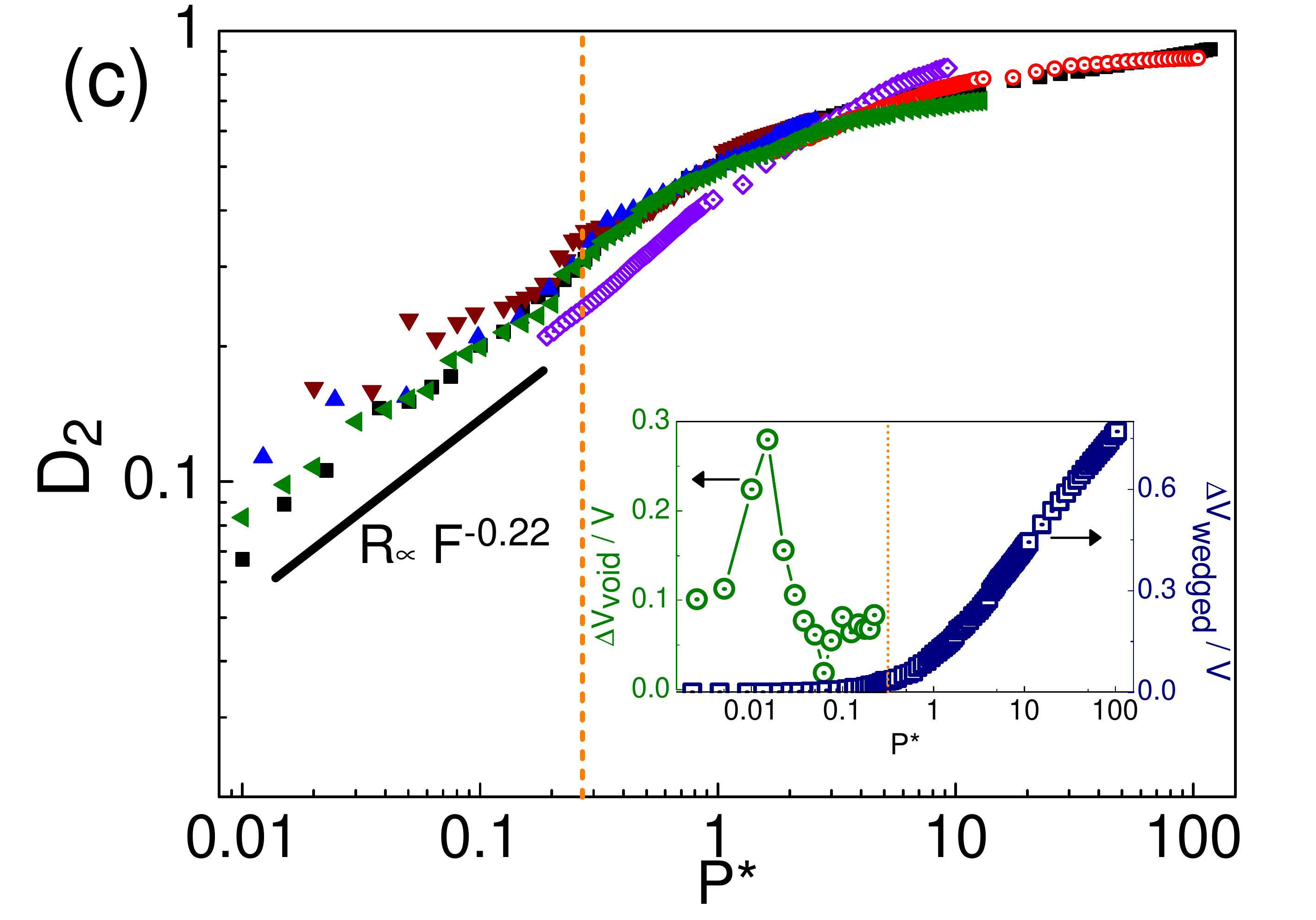}
\label{F_Density_mod2}
\end{minipage}
}
\caption  {(a) Schematic plot for a (red) bead partially wedged in the interior (the grey prism) of a triangular lattice of (blue) beads. (b) Void (in black color) of a crumpled ball
(c) Dimensionless density $D_2$ vs. dimensionless pressure $P^\star$. Data are replotted from Fig.\ref{fig_P_Density}. The power-law region shrinks from $P^\star\le 1.8$ to 0.28, and $\alpha$ decreases from 0.25 to about 0.22. Inset shows that $\Delta V_{\rm void}/V$ and $\Delta V_{\rm bead}/V$ dominate the small and large pressure region, respectively, for the case of a $k_B=1000$ elastic sheet.
}\label{number} 
\end{figure}

The reason leading to the discrepancy is mainly due to the coarse-grained triangular lattice 
model used here. The actual volume of sheet is underestimated in the numerator of $D_1$, which should be $\pi R_0^2 h$ where $h$ is the thickness of sheet. Also, the volume of the crumpled sheet in the denominator of $D_1$ is overestimated. A correct value can be calculated from simulations which is smaller than the volume of spherical boundary $V=4\pi R_{\rm wall}^3/3$ by the amount of void $\Delta V_{\rm void}$ (the black volume in Fig.~\ref{number}(b)). However, since a real paper can not cut into itself, the beads that are increasingly wedged in the interstitial holes of the triangular lattices as crumpling progresses, as sketched in Fig.~\ref{number}(a), should not be allowed. Had this constraint been imposed, the crumpled ball size would swell by approximately the total amount of the wedged volume $\Delta V_{\rm wedged}$.

Therefore, an appropriate dimensionless density should be defined as 
$D_2=(\pi R_0^2 h)/(V-\Delta V_{\rm void}+\Delta V_{\rm wedged})$.
Figure \ref{number}(c) presents the plot $D_2$ vs.~$P^\star$ for the different sets of 
simulation data together with the experimental ones, where 
the ratios for the volume differences, $\Delta V_{\rm void}/V$ and $\Delta V_{\rm wedged}/V$ 
are also shown in the inset for illustration.
We observe that the sets of simulation data are now collapsed with the experimental ones, while retaining the scaling relation. In the mean time, the power-law region shrinks and the exponent reduces to around 0.22. The latter is not hard to understand because $\Delta V_{\rm void}/V$ in the inset of Fig.\ref{number}(c) dominates at small density and raises $D_1$, while $\Delta V_{\rm wedged}/V$ rules at large density to reduce $D_1$. Both effects combine to level off the line that represents the power law in Fig.\ref{fig_P_Density}.

Please notice that experimentalists usually focus on the response in a relatively 
large density region where the power-law relation is not so well-established, 
as shown in Fig.~\ref{number} for Al and HDPE. 
It explains why $\alpha$ obtained from experiments\cite{lin,bai} is usually smaller and non-universal.
This is because there exists a lower limit to apply external pressure $P$ 
in crumpling experiments\cite{lin,bai} where high-pressure nitrogen gas was used to provide the ambient pressure. 
Since the pressure is balanced, the sheet needs to be pre-crumpled and packed by a PVC wrap
with the wrap interior maintained at atmospheric pressure via a PE tube 
connecting to the outside of the pressure chamber. 
This procedure sets a lower value to $P$ and, therefore, it is not easy for experimentalists to
investigate crumpling in the low-pressure region.

\begin{figure}[!h]
  \centering
  \includegraphics[width=7cm]{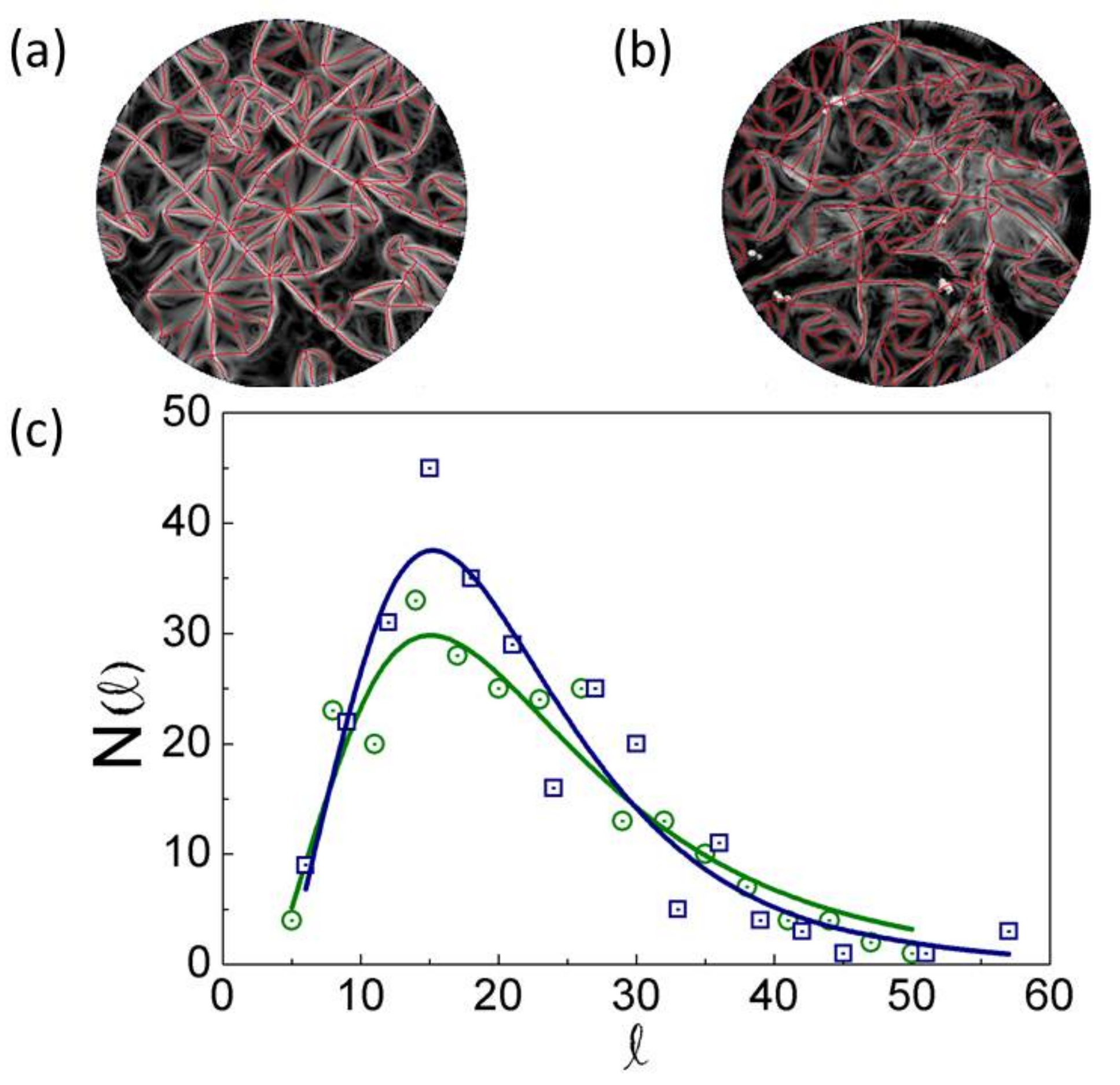} 
  \caption
  {(a) Network of ridges on a crumpled elastic sheet of  $k_B=1000$ at $D_2=0.339$.
  (b) Same as (a) but on a plastic sheet.  (c) $N(\ell)$ as a function of $\ell$.
  The distribution is fitted by $N(\ell)=A/(\ell \sqrt{B})\exp\{-[ln(\ell/\ell_0)/B]^2\}$ with 
  $(A,B,\ell_0)=(429.6, 0.66, 20.5)$ for elastic sheets (circles) and $(449.9, 0.48, 19.0)$ for
   plastic sheets (squares). 
  }\label{fig3}
\end{figure}

\begin{figure}[!h]
  \centering
  \includegraphics[width=7cm]{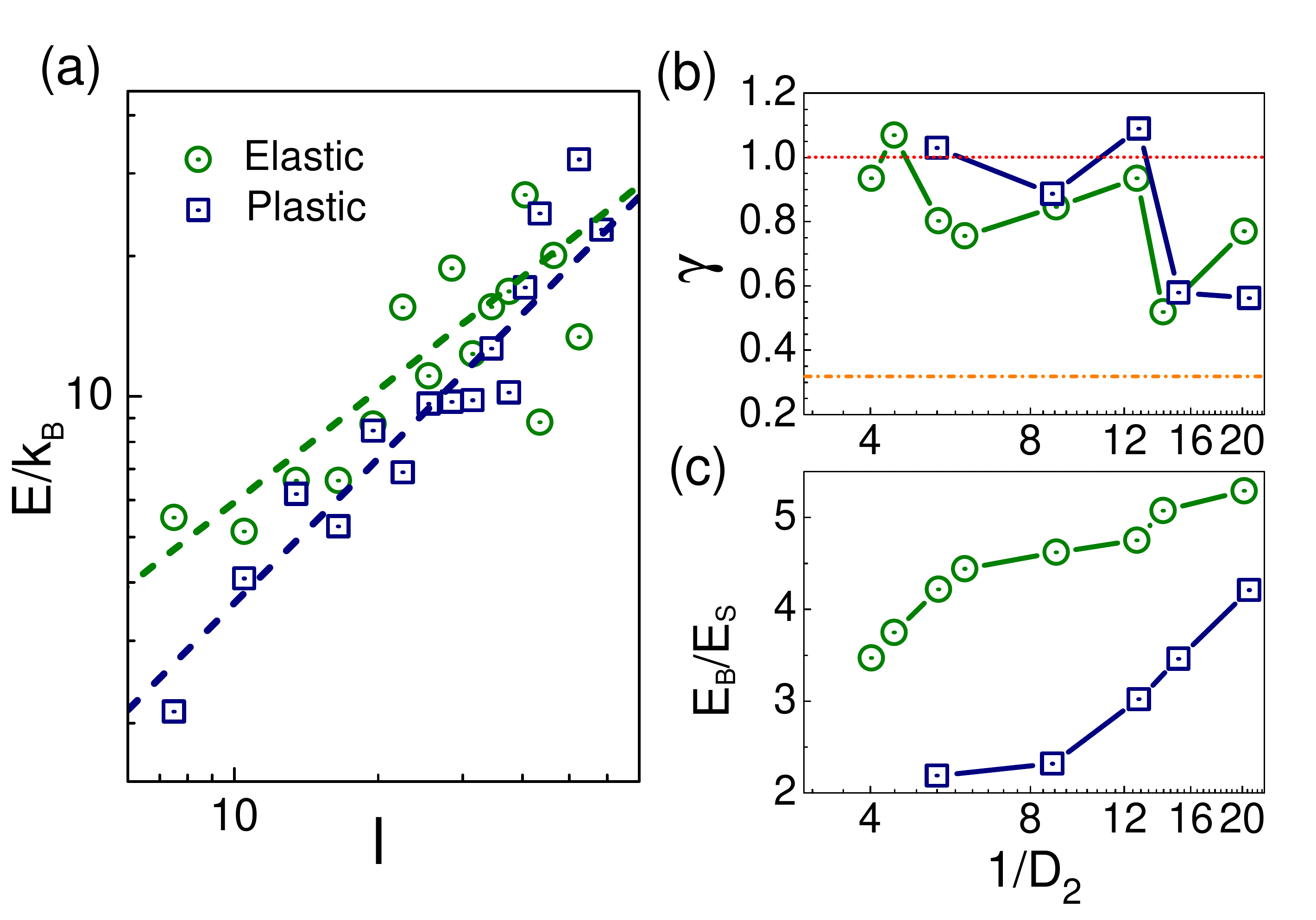}  
  \caption
  {(a) Averaged storing energy $E(\ell)$ in a ridge as a function of $\ell$ 
  at $D_2=0.339$ and $k_B=1000$. Least square fit yields $\gamma =0.802$ and 1.029 for the elastic and plastic sheets.    (b) $\gamma$ vs.~$1/D_2$ (c) $E_B/E_S$ vs.~$1/D_2$   
}\label{fig4} 
\end{figure}

	It is known that the single-ridge approximation\cite{ridge,witten,wood} predicts a scaling law for the 
energy $E$ to create a folding ridge on a sheet as a one-third power of 
the ridge length $\ell$. 
If $E$ is partitioned into the bending energy $E_B$ and the stretching energy $E_S$, 
the ratio of the two energies is shown to be $E_B/E_S=5$.  
Although the former prediction has been checked by simulations for a solitary ridge\cite{tallinen2}, 
the two predictions have not yet been verified in a realistic case of crumpling 
where multiple ridges are formed and interact between each other.
Since the ridges are not formed in an equilibrium state, a full relaxation of stress is not expected, particularly when the sheet is crumpled into a very tight space.

We apply watershed algorithm to determine the folding ridges of sheet. 
The data are improved manually by connecting broken segments of the ridges obtained 
(see Fig.\ref{fig3}(a),(b) for example).
We observe that the distribution of ridge length follows a log-normal distribution 
as shown in Fig.\ref{fig3}, which agrees with theoretical predictions\cite{ridge,wood}, 
experiments\cite{blair,andresen} and other simulations\cite{vliegenthart}. 
Moreover, we calculate the energy associated to each ridged generated.
The mean energy $E(\ell)$ exhibits a power-law dependence $\ell^\gamma$ 
on the ridge length $\ell$, as shown in Fig.\ref{fig4}(a).
The exponent $\gamma$ is then plotted in Fig.\ref{fig4}(b) as a function of density $D_2$. 
We find that the predicted value $1/3$ of Witten's single-ridge approximation is recovered 
in our simulations at low densities.  
As the density increases, $\gamma$ rises and saturates asymptotically to 1.
In the mean while, the ratio $E_B/E_S$ drops from the single-ridge value 5 to about 2 (see Fig.\ref{fig4}(c)). Therefore, the weight of energy gradually shifts from $E_B$ to $E_S$ 
as $D_2$ increases, whereas both energies are promoted by the ridge-ridge interactions.
Since the dependence of $E_B/E_S$ on $D_2$ cannot be predicted by the single-ridge 
approximation, our results show that the 
ridge-ridge interactions play an important role on the energy storage of sheet. 

\begin{figure}[!h]
  \centering
  \includegraphics[width=7cm]{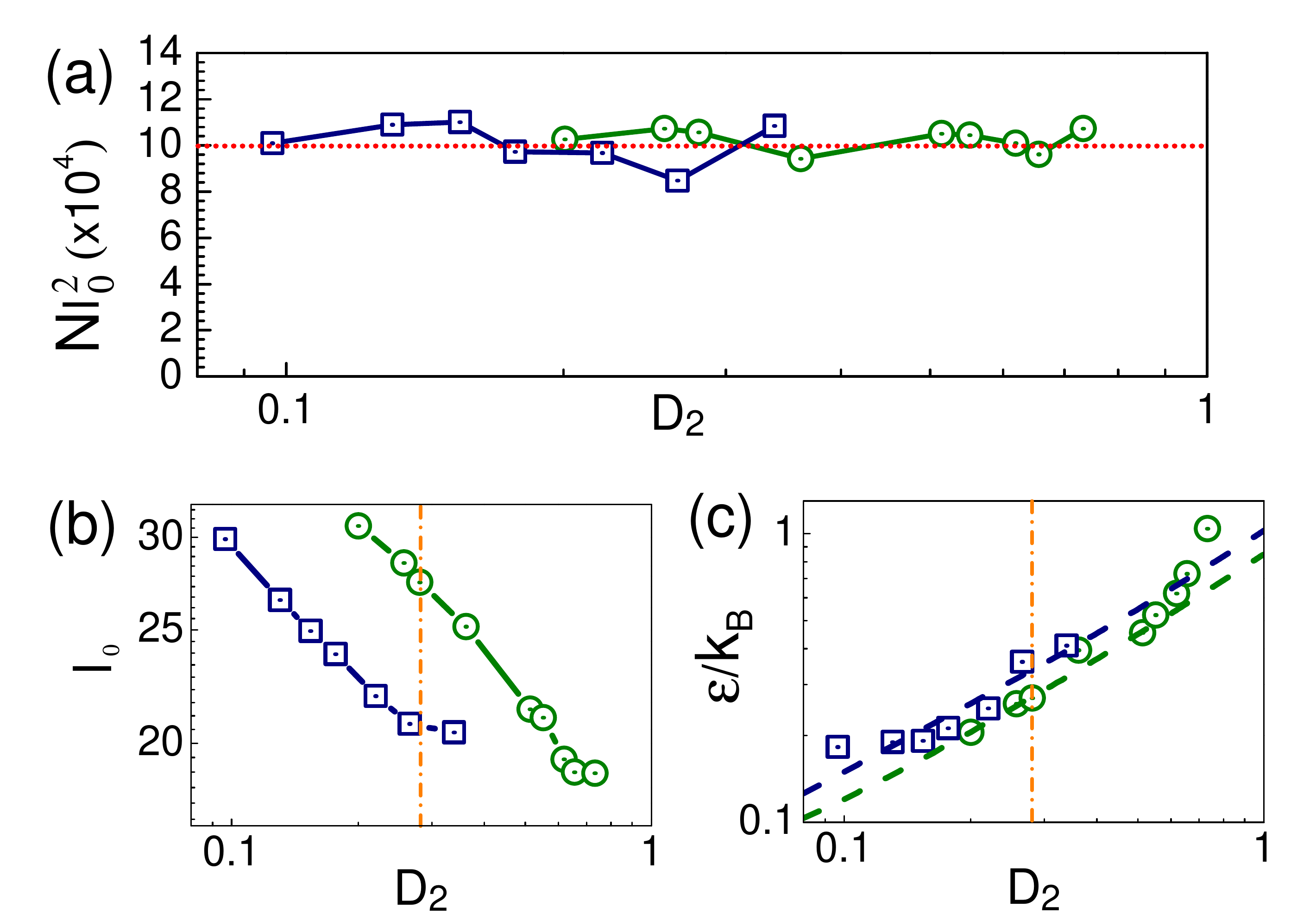} 
  \caption
  {(a) $N \ell_0^2$ vs. $D_2$ for an elastic sheet with $k_B=1000$.
  (b) $\ell_0$ vs.~$D_2$. The power-law exponents are  -0.292 and -0.307, respectively, for elastic (circles) and plastic (squares) sheets.
  (c) Power law dependence of $\varepsilon$ on $D_2$. The exponents are 0.788 and 0.829, respectively.
}\label{fig5} 
\end{figure}

Having seen how the ridge-ridge interactions revise the energy scalings, we now study
some properties that turn out to obey the prediction of a simple mean-field theory. 
As shown in Fig.~\ref{fig5}(a), the ridge number $N$ multiplies the square of 
the mean ridge length $\ell_0$ is essentially constant over the whole range of density $D_2$.
Suppose that the ridges are arranged in a square lattice, the area of this checkboard is approximately $N\ell_0^2/2$, which is consistent with the sheet area $\pi R_0^2=5.3\times 10^4$. 
Thus, the fact that $N\ell_0^2$ does not change with $D_2$ can be explained within a mean-field picture.
Figure \ref{fig5}(b) shows that $\ell_0\sim 1/D_2^{1/3}$ or equivalently $\ell_0\sim R$. This result can be explained simply through a dimensional analysis:
$\ell_0$ scales merely with the length scale $R$ of system
and not with $k_B$ and $k_S$,
because ridges appear also on a crumpled sheet even in the condition $k_B=k_S=0$.

Furthermore, the average energy per unit ridge length follows a power law
$\varepsilon \sim D_2^{0.788\sim 0.829}$ in the low-density region $D_2<0.28$, 
as shown in Fig.~\ref{fig5}(c), which is also the region where $R \sim F^{-\alpha}$.
It implies that the total energy stored in the ridges, $N\ell_0\varepsilon$,
scales as $\ell_0^{-1} R^{-3\times0.788 \sim -3\times 0.829}$, or $R^{-3.364\sim -3.487 }$.
Because of energy conservation, the total work $W$ done on the sheet during crumpling is
equal to the increase of internal energy. 
$W$ can be calculated by $\int^R_{R_0} F \cdot dR'$, and 
yields a scaling $R^{(-1/\alpha)+1}$ as $R \ll R_0$, since $F \sim R^{-1/\alpha}$.
For an exponent $\alpha\approx 0.22$ as obtained in Fig.\ref{number}(c), 
the scaling reads as $W\sim R^{-3.54}$, which is in good agreement with 
the scaling relation for $N \ell_0 \varepsilon$ here.

In conclusion, Molecular Dynamics Simulation has been employed to study the effect of ridge-ridge interactions on a crumpling of self-avoiding sheet. Energy scaling based on the single-ridge approximation is valid only at low densities. Revision is needed as the ridge-ridge interaction intensifies in the high-density region. We manage to conciliate the different exponents $\alpha$ between simulations and experiments. We investigate how the average ridge number, 
the average ridge length, and the average storing energy per unit length vary with the density. 
The relations are found to be consistent with the mean-field predictions, in spite of the strong ridge-ridge interactions. Particularly, we extend our simulations to the high-pressure region and confirm the existence of a general scaling relation which 
is valid for the whole range of densities.

We thank the National Science Council in Taiwan for financial support, Yenchih Lin for helpful discussions, and the Physics Division of National Center for Theoretical Sciences in Hsinchu for hospitality.


\end{document}